\documentclass[12pt,preprint]{aastex63}
\usepackage{multirow}

\newcommand{\kms}{\rm km~s^{-1}}
\newcommand{\kmsmpc}{\rm km~s^{-1}~Mpc^{-1}}
\newcommand{\dn}{\rm D_{n}4000}
\newcommand{\msol}{\rm M_{\odot}}
\newcommand{\mchar}{\rm M_{200}}
\newcommand{\rchar}{\rm R_{200}}
\newcommand{\rcl}{\rm R_{cl}}

\begin{document}

\title{A Spectroscopic View of the JWST/GTO Strong Lensing Cluster A1489}

\author{Kenneth J. Rines}
\affiliation{Department of Physics and Astronomy, Western Washington University, Bellingham, WA 98225, USA}

\author{Jubee Sohn}
\affiliation{Smithsonian Astrophysical Observatory, 60 Garden Street, Cambridge, MA 02138, USA}

\author{Margaret J. Geller}
\affiliation{Smithsonian Astrophysical Observatory, 60 Garden Street, Cambridge, MA 02138, USA}

\author{Antonaldo Diaferio}
\affiliation{Universit{\`a} di Torino, Dipartimento di Fisica, Torino, Italy}
\affiliation{Istituto Nazionale di Fisica Nucleare (INFN), Sezione di Torino, via P. Giuria 1, I-10125 Torino, Italy}

\email{rinesk@wwu.edu}

\begin{abstract}
We discuss a spectroscopic survey of the strong lensing cluster A1489 that includes redshifts for 195 cluster members along with central velocity dispersions for 188 cluster members. The caustic technique applied to the redshift survey gives the dynamical parameters $\mchar = (1.25~\pm~0.09) \times 10^{15}~\msol$, $\rchar = 1.97~\pm~{0.05}$ Mpc, and a cluster line-of sight velocity dispersion $1150~\pm~{72}~\kms$ within $\rchar$. These parameters are  very similar to those of other strong lensing systems with comparably large Einstein radii. We use the spectroscopy and deep photometry to demonstrate that A1489 is probably  dynamically active; its four BCGs have remarkably different rest frame radial velocities. Like other massive strong lensing clusters, the velocity dispersion function for members of A1489 shows an excess for dispersions $\gtrsim~250~\kms$. The central dispersions also provide enhanced constraints on future lensing models.
\end{abstract}

\section{Introduction}

Strong lensing clusters of galaxies enable unique cosmological insights. They provide some of the tightest constraints on  inner cluster mass profile (e.g., \citealp{Bartelmann10, Kneib11}) and highly magnified views of distant young galaxies (some recent examples include \citealp{Zheng12, Coe13, Laporte17, Salmon20}). Strong lensing clusters are a remarkable window on the high redshift universe.

Surveys with HST have often focused on clusters with potentially large Einstein radii because of their impressive power for exploration of the high redshift universe. These systems were generally selected based on their X-ray or SZ properties \citep{Planck16,Repp16}. To enlarge the sample of these amazing systems, investigators have extended the search to large optical catalogs like redMaPPer \citep{Rykoff14, Rykoff16}. Some of the richest and most centrally concentrated of the optically identified clusters are strong lensing systems. One of the optically identified strong lensing systems, A1489 ($z = 0.351$), is also a JWST/GTO target.

Here we discuss a spectroscopic survey of A1489 that includes redshifts for 183 members within 9 Mpc of the cluster center. We also provide central stellar velocity dispersions; these observations promise enhancement of constraints on lensing models (e.g., \citealp{Monna17, Bergamini19, Granata21, Acebron21}).

Optical spectroscopy of strong lensing clusters provides physical insights into these systems that complement strong lensing. For example, large samples of cluster members with redshifts provide dynamical mass estimates to large radius (e.g., \citealp{Rines06, Biviano13, Rines13, Sohn17a}) and they can  provide a measure of infall onto the system \citep{Pizzardo21}. Strong lensing clusters may show evidence of dynamical activity. Substructures in the central region and  at large radius can be resolved with an intensive spectroscopic survey (e.g., \citealp{Czoske02, Balestra16, Mahler18, Richard21}).

We begin with a review of the history of A1489 (Section \ref{sec:history}). We then describe the redshift survey and underlying and/or related data (Section \ref{sec:data}). We discuss the members of A1489 in Section \ref{sec:member} and compare their velocity dispersion function with other lensing systems in Section \ref{sec:vdf}. We derive the dynamical parameters of A1489 in Section \ref{sec:dmass} and discuss the apparent departure from equilibrium in Section \ref{sec:complex}. We also review the use of central stellar velocity dispersions as constraints on lensing models in Section \ref{sec:lensing}. We conclude in Section \ref{sec:conclusion}. We use the standard $\Lambda$CDM cosmology with $H_{0} = 70~\kmsmpc$, $\Omega_{m} = 0.3$, $\Omega_{\Lambda} = 0.7$, $\Omega_{k} = 0.0$ throughout.

\section{A Brief History of A1489}\label{sec:history}

\begin{figure} [ht!]
\centering
\includegraphics[scale=0.48]{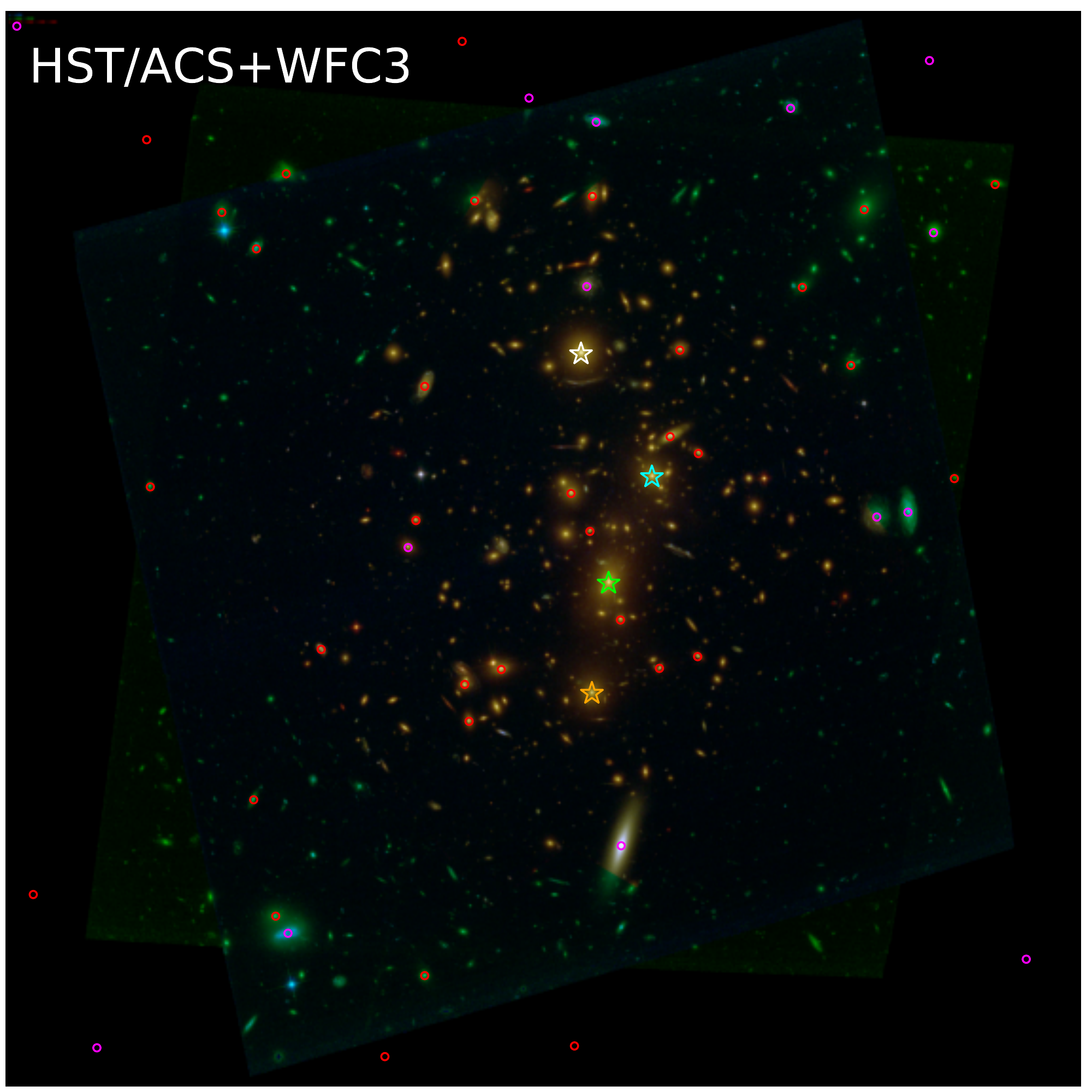}%{jsohn_figure/A1489_hst.pdf}
\caption{Color image of A1489 obtained from HST ACS and WFC3 (courtesy of Dan Coe). Red circles show spectroscopically identified cluster members, and magenta circles are the non-member galaxies with spectroscopic redshifts. The four colored stars mark the 4 bright cluster members denoted A (white), B (green), C (cyan), and D (yellow) in \citet{Zitrin20}.}
\label{hst}
\end{figure}

Abell 1489 gets its original name from inclusion in the Abell catalog \citep{Abell58}. Abell identified clusters from the Palomar Sky Survey. The typical redshift for a cluster in the catalog is $z \leq 0.2$. A1489 at $z = 0.351$ is probably included because of its richness and because of the presence of several bright galaxies in its core (Figure \ref{hst}). In fact, A1489 (RMJ121218.5+273255.1) is among the 0.1\% richest clusters in the optically selected redMaPPer catalog \citep{Rykoff14,Rykoff16}. This catalog, based on Sloan Digital Sky Survey (SDSS) DR8 photometry, has a median depth of $z \sim 0.35$.

A1489 is also an extended X-ray source labeled CL1212+2733 \citep{Burenin07}. The cluster was initially included in the Chandra Cluster Cosmology Project \citep{Vikhlinin09}. The fiducial X-ray mass is $\sim 10^{15}$ M$_\odot$ commensurate with its richness. The \citet{Planck16} also detect the cluster.

%========================================
%========================================

The JWST Medium-Deep Fields Program (P.I. Rogier Windhorst) focuses on several lensing clusters, some of which were optically selected, that form or are likely to form a large strong gravitational lens. A1489 is part of this survey. Its Einstein radius is $\Theta_E = 32 \pm 3 \arcsec~$ for a source redshift $z_{S} = 2$ \citep{Zitrin20}. 

As a foundation for the JWST observations, \citet{Zitrin20} acquired deep HST  and Gemini Multi-Object Spectrograph (GMOS) on Gemini-N imaging of the cluster as a basis for a detailed strong lensing model. \citet{Zitrin20} estimated photometric redshifts for candidate lensing arcs. They identified potential cluster members based on the photometric red sequence. 

Figure \ref{hst} shows the HST image of A1489 (\citealp{Zitrin20}, image courtesy of Dan Coe and Anton Koekemoer\footnote{\url{https://www.stsci.edu/~koekemoer/zitrin/RMJ1212}}). In the image, bright red circles mark galaxies with spectroscopy (included in this paper) that are cluster members. Magenta circles mark spectroscopically identified non-members. We describe the spectroscopy in Section \ref{sec:redshift} and the membership determination in Section \ref{sec:member}.

The colored stars in Figure \ref{hst} indicate the brightest four cluster members in the $r-$band. These objects are identical to the galaxies denoted A (white), B (cyan), C (green), and D (yellow) in \citet{Zitrin20}. For efficiency, we refer to these galaxies as BCGs hereafter. Table \ref{tab:a1489} lists the basic properties of A1489 that we derive in this paper. The Table includes the total number of galaxies in the redshift survey ($N_{spec}$) and the number of spectroscopic cluster members ($N_{member}$). Table \ref{tab:a1489} also lists the position, mean redshift, and a characteristic radius, mass, and velocity dispersion derived from the redshift survey.

%=================================
%Table \ref{tab:a1489}
%=================================
\begin{deluxetable}{cccccccc}
\label{tab:a1489}
\tablecaption{Basic properties of A1489}
\tablecolumns{8}
\tabletypesize{\scriptsize}
\tablewidth{0pt}
\tablehead{
\colhead{R.A.} & \colhead{Decl.} & \colhead{z} & \colhead{$R_{200}$} & \colhead{$M_{200}$} & \colhead{$\sigma_{cl}$} & \colhead{$N_{spec}$} & \colhead{$N_{member}$} \\
\colhead{(deg)} & \colhead{(deg)} & \colhead{} & \colhead{(Mpc)} & \colhead{($M_{\odot}$)} & \colhead{(km s$^{-1}$)} & \colhead{} & \colhead{} }
\startdata
183.075770 & 27.552620 & 0.350870 & $1.97 \pm 0.05$ & $(1.25 \pm 0.09) \times 10^{15}$ & $1150 \pm 72$ & 1010 & 195
\enddata 
\end{deluxetable}
%=================================

\section{The Data} \label{sec:data}

In spite of its selection as a JWST/GTO target, there are relatively few spectroscopic measurements for A1489 cluster members. Here we discuss a redshift survey carried out in 2007 as part of an exploration of 10 clusters in the Chandra Cluster Cosmology Project. We describe the photometric basis for the survey along with subsequent photometry from DECaLS \citep{Dey19} (Section \ref{sec:photometry}), redshift measurements with Hectospec on the MMT (Section \ref{sec:redshift}), derivation of the spectral index $\dn$ and the stellar velocity dispersion from the MMT data (Section \ref{vdisp}) and computation of stellar masses (Section \ref{smass}).

\subsection{Photometry} \label{sec:photometry}

We based the original selection of spectroscopic targets on the SDSS DR6 \citep{SDSSDR6} available at the time the observations were made. The targets have $r_{cmodel} \leq 20.8$ (composite model magnitudes are a blend of the best-fit exponential and de Vaucouleurs model fits to the SDSS photometry). For efficient acquisition of reasonable quality spectra we include a central surface brightness limit, $\mu_r \leq 22.9$ within the half-light Petrosian radius. In the outer region where the contrast between the cluster members and the foreground/background decreases substantially, we target only objects with $g-r > 1.3$ to eliminate most galaxies with $z \lesssim 0.3$.

Recent DECaLS (the Dark Energy Camera Legacy Survey) photometry \citep{Dey19} is a substantial improvement over the SDSS. We thus match the survey catalog to DECaLS\footnote{One DECaLS object at (R.A., Decl.) = (183.197929,27.543698) is an unresolved pair of galaxies; we replace it with SDSS photometry.}. Figure \ref{complete} shows the completeness of the survey within $R < \rchar = 1.97$ Mpc as a function of DECaLS $r-$band apparent magnitude\footnote{$\rchar$ is the characteristic radius enclosing a mean density 200 times the critical density of the universe at the cluster redshift.}.

\subsection{Redshifts}\label{sec:redshift}

In 2007 May and June, we acquired spectra with Hectospec, a 300-fiber instrument with a $1^\circ$ field mounted on the MMT 6.5m telescope \citep{Fabricant05}. A single Hectospec observation typically acquires 250 target spectra. Each Hectospec spectrum, obtained through a 270 mm$^{-1}$ grating, covers the wavelength range 3700 -- 9100 {\rm \AA~} with an average resolution of 6.2 {\rm \AA}. We observed A1489 with four configurations of Hectospec, three configurations of 2700s each in 2007 May, and a fourth configuration in 2007 June (under sub-optimal observing conditions) for 4800s. We re-observed some of the targets from the 2007 May configurations because the spectra did not yield reliable redshifts.

We reduced the Hectospec spectra using the standard HSRED v2.0 package\footnote{\url{http://mmto.org/~rcool/hsred/}}. We measured the redshift using cross-correlation (RVSAO, \citealp{Kurtz98}). RVSAO yields the cross-correlation score $R_{XC}$. We visually inspected all spectra to confirm the reliability of the redshift. A score of $R_{XC} > 5$ indicates a reliable redshift; we include some galaxies with $R_{XC} < 5$ when visual inspection of the spectrum reveals multiple obvious absorption and/or emission lines. Reduced 1D spectra are available from the MMT archive\footnote{\url{https://oirsa.cfa.harvard.edu/archive/search/}}. 

Table \ref{tab:redshift} lists 1010 objects with spectroscopy in the A1489 field. We obtained spectroscopy for 898 objects with Hectospec and we collected 213 redshifts from SDSS. There are 101 overlaps between Hectospec and the SDSS; we include only the Hectospec redshift in these cases. Table \ref{tab:redshift} includes the SDSS ObjID, the R.A., Decl., the redshift, the DECaLS r-band magnitude, DECaLS $(g-r)$, the spectral indicator D$_n$4000, the stellar mass, $M_{*}$, the central velocity dispersion $\sigma_{*, 3 kpc}$, and a cluster membership indicator. 

%=================================
%Table \ref{tab:redshift}
%=================================
\begin{deluxetable}{lcccccccccc}
\label{tab:redshift}
\tablecaption{A1489 Redshift Survey}
\tablecolumns{11}
\tabletypesize{\scriptsize}
\tablewidth{0pt}
\tablehead{
\multirow{2}{*}{SDSS Object ID} & \colhead{R.A.} & \colhead{Decl.} & \multirow{2}{*}{z} & \colhead{$r$} & \multirow{2}{*}{$g-r$} & \multirow{2}{*}{$\dn$} & \colhead{$\log M_{*}$} & \colhead{$\sigma_{*, 3 kpc}$} & \multirow{2}{*}{Member} & \multirow{2}{*}{SDSS} \\
                                & \colhead{(deg)} & \colhead{(deg)} & & \colhead{(mag)} & & & \colhead{($\msol$} & \colhead{($\kms$)} & & }

\startdata
1237667322719240304 & 182.646527 & 27.241936 & $ 0.342316 \pm 0.000157$ & 19.82 & 1.39 & $1.33 \pm 0.04$ & $ 10.37 \pm _{0.14}^{0.19}$ & $136 \pm 21$ & N & N \\
1237667322719306155 & 182.839128 & 27.128418 & $ 0.273392 \pm 0.000158$ & 20.08 & 1.34 & $1.60 \pm 0.06$ & $ 10.37 \pm _{0.12}^{0.16}$ & $187 \pm 30$ & N & N \\
1237667322719306265 & 182.877989 & 27.145353 & $ 0.487004 \pm 0.000164$ & 20.09 & 1.77 & $1.65 \pm 0.08$ & $ 11.27 \pm _{0.11}^{0.09}$ & $182 \pm 26$ & N & Y \\
1237667322719306104 & 182.818271 & 27.145142 & $ 0.385302 \pm 0.000118$ & 20.01 & 1.34 & $1.28 \pm 0.09$ & $ 10.75 \pm _{0.18}^{0.17}$ & $ 76 \pm 55$ & N & N \\
1237667322719240591 & 182.714070 & 27.181036 & $ 0.318616 \pm 0.000110$ & 18.82 & 1.54 & $1.84 \pm 0.05$ & $ 10.99 \pm _{0.14}^{0.10}$ & $260 \pm 23$ & N & N \\
1237667322719240353 & 182.766365 & 27.200521 & $ 0.559170 \pm 0.000098$ & 19.83 & 1.78 & $1.58 \pm 0.07$ & \nodata & $233 \pm  24$ & N & Y \\
1237667322719240695 & 182.774884 & 27.166615 & $ 0.262940 \pm 0.000101$ & 19.13 & 1.27 & $1.76 \pm 0.06$ & $ 10.61 \pm _{0.19}^{0.09}$ & $161 \pm 18$ & N & N \\
1237667322719305927 & 182.788763 & 27.197877 & $ 0.344549 \pm 0.000086$ & 20.21 & 1.66 & $1.79 \pm 0.05$ & $ 10.75 \pm _{0.14}^{0.13}$ & $171 \pm 14$ & N & N 
\enddata 
\end{deluxetable}
%=================================

%=================================
%Table \ref{tab:bcg}
%=================================
\begin{deluxetable}{cccccccccc}
\label{tab:bcg}
\tablecaption{A1489 BCGs}
\tablecolumns{10}
\tabletypesize{\scriptsize}
\tablewidth{0pt}
\tablehead{
\multirow{2}{*}{BCG ID} & \colhead{R.A.} & \colhead{Decl.} & \multirow{2}{*}{z} & \colhead{$R_{proj}$} & \colhead{$\Delta V$\tablenotemark{*}} & \multirow{2}{*}{$\dn$} & \colhead{$\log M_{*}$} & \colhead{$\sigma_{*, 3 kpc}$} & \multirow{2}{*}{Symbol color} \\
                        & \colhead{(deg)} & \colhead{(deg)} & 
                   & \colhead{(Mpc)} & \colhead{($\kms$)} & & 
\colhead{$\msol$} & \colhead{($\kms$)} & }
\startdata
A & 183.079172 & 27.564635 & 0.357564 & 0.221 & 1486  & 1.97 & 11.60 &  325 & White \\
B & 183.077005 & 27.548645 & 0.350695 & 0.073 & -39   & 2.06 & 11.35 &  337 & Green \\
C & 183.073596 & 27.556057 & 0.356577 & 0.070 & 1267  & 1.81 & 11.01 &  257 & Cyan  \\
D & 183.078317 & 27.540965 & 0.345509 & 0.211 & -1190 & 1.94 & 11.46 &  209 & Orange 
\enddata 
\tablenotetext{*}{$\Delta V = c(z_{BCG} - z_{cl}) / (1 + z_{cl})$}
\end{deluxetable}
%=================================

There is a small offset between the Hectospec and SDSS/BOSS redshifts estimated from 981 overlapping objects in HectoMAP (i.e., $\Delta c(z_{SDSS/BOSS} - z_{Hecto}) / (1 + z_{Hecto})$) $\sim 39~\kms$ \citep{Sohn21}, comparable with the typical uncertainty in the Hectospec redshift ($\sim 40~\kms$). The offset between the SDSS and Hectospec redshifts is irrelevant for the analysis we undertake here. Among the members of the cluster, only 6 have redshifts from SDSS alone.

Table \ref{tab:bcg} highlights the properties of the 4 BCGs denoted by colored stars in Figure \ref{hst}. The Table includes the IDs from \citet{Zitrin20} in column (1) followed by the the right ascension, declination, redshift, radial offset, and rest-frame velocity offset. The BCG parameters derived in the following sections include the spectral indicator $\dn$, the stellar mass, and the line-of-sight stellar velocity dispersion within a 3 kpc aperture. The final column of the Table gives the color that denotes each of these BCGs in Figure \ref{hst} and in the figures that follow.

\begin{figure}[ht!]
\centering
\includegraphics[scale=0.38]{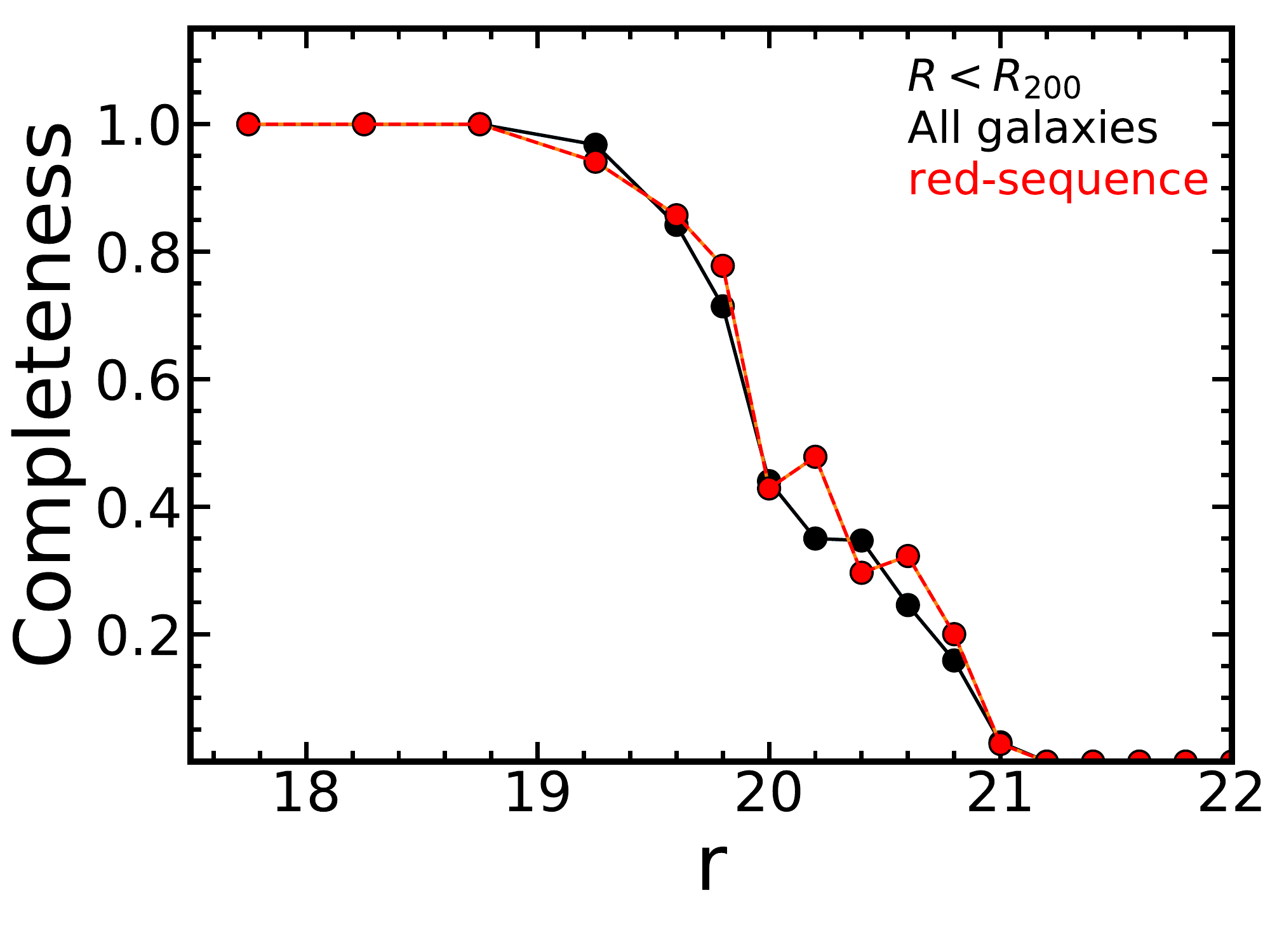}
%{jsohn_figure/A1489_completeness_r.pdf}
\caption{Spectroscopic survey completeness as a function of $r-$band magnitude. Black and red symbols show the survey completeness for all the galaxies and for the galaxies on the red-sequence. }
\label{complete}
\end{figure}

Figure \ref{complete} shows the differential completeness of the entire redshift survey for DECaLS $r$-band magnitude ($\rcl < \rchar$). Red symbols indicate the completeness for galaxies on the red sequence (see below) and black points indicate the completeness regardless of color. The integrated completeness exceeds 80\% for $r < 20$ mag and drops steeply for fainter objects.

\begin{figure}[ht!]
\centering
\includegraphics[scale=0.38]{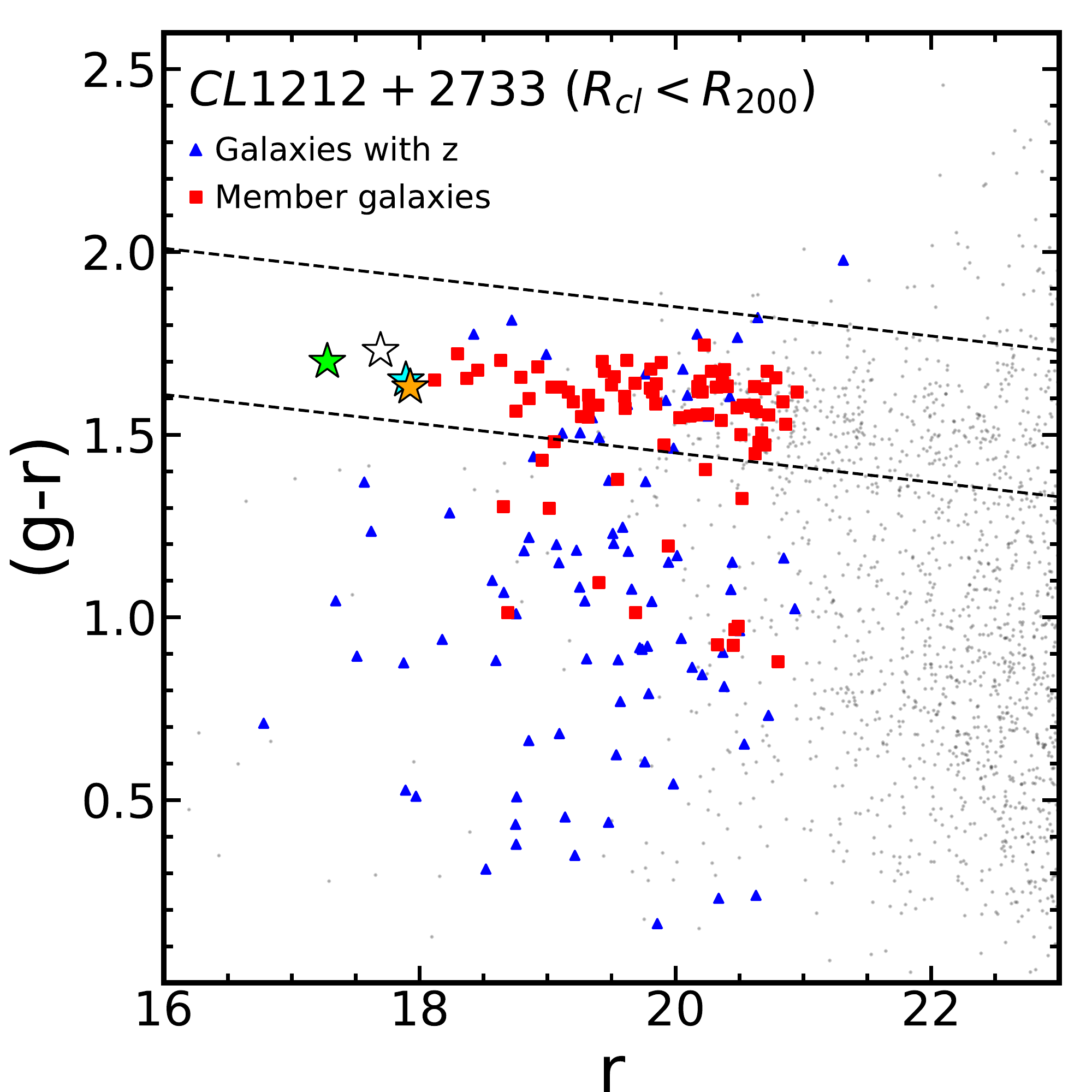}
%{jsohn_figure/A1489_cmd.pdf}
\caption{Color magnitude diagram for galaxies within $\rcl < \rchar$ Mpc of the A1489 (CL1212+2733) center. Red squares show cluster members; blue triangles show non-member galaxies with spectroscopy. Background black dots show DECaLS galaxies within the field ($r < 23$ mag). The four colored stars denote the BCGs as in Figure \ref{hst}.}
\label{cmd}
\end{figure}

Figure \ref{cmd} shows the color magnitude diagram for galaxies with measured redshifts that lie within $\rchar$ of the cluster center. We define the red sequence following \citet{Rines13}. In $(g-r)$ versus $r$ color magnitude diagram, we assume a  red-sequence slope, $-0.04$. We determine the intercept based on linear fitting of the member galaxies. We define galaxies within $\pm 0.2$ (the two dashed lines) of the red-sequence as red-sequence members. Here again the four colored stars represent the 4 BCGs; they are the brightest galaxies on the red sequence in the r-band.

%========================================
%========================================

%========================================
%========================================

\begin{figure}[ht!]
\centering
\includegraphics[scale=0.38]{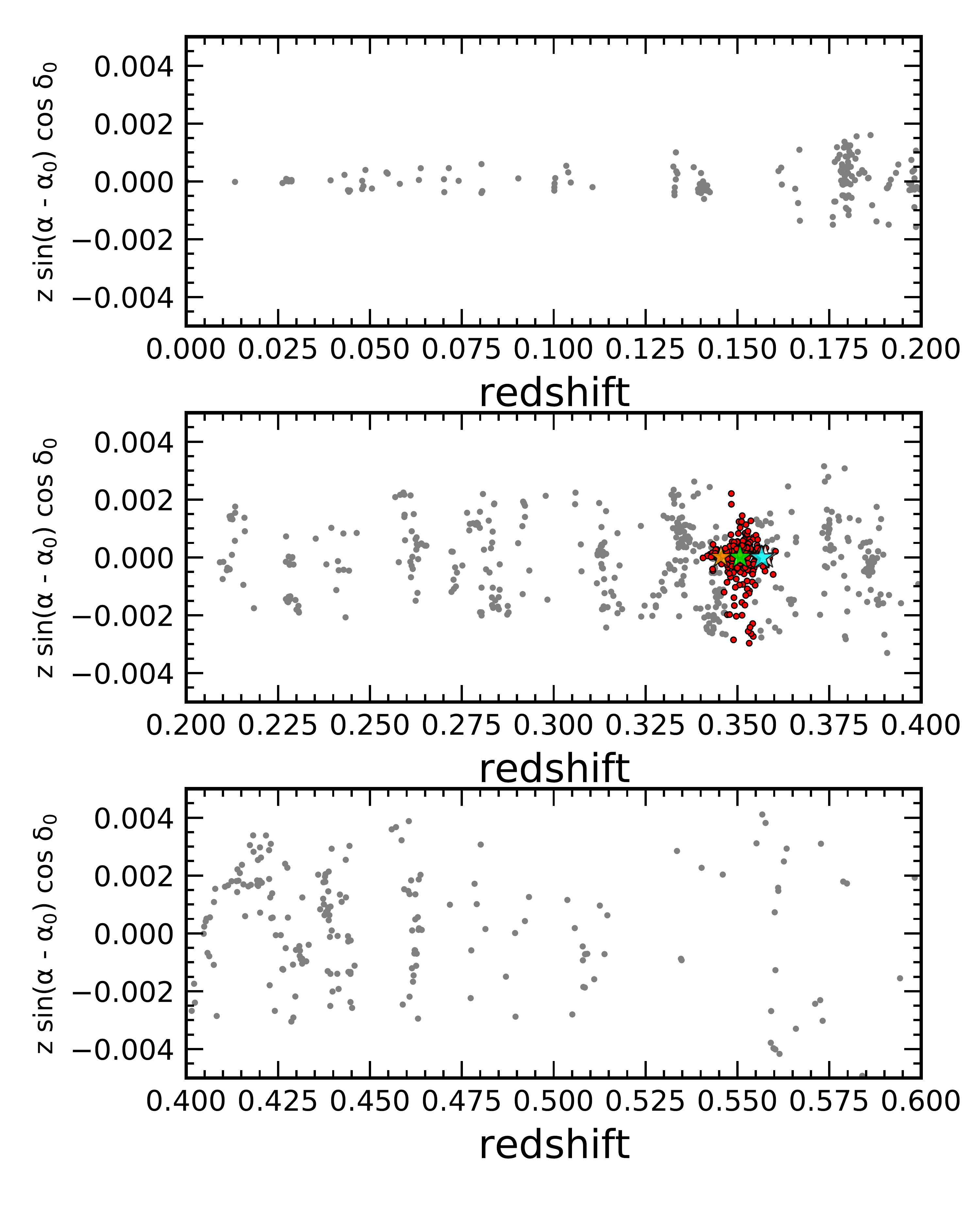}
%{jsohn_figure/A1489_cone.pdf}
\caption{Cone diagram for the A1489 field. Red circles show spectroscopically identified A1489 members. Colored stars denote the 4 BCGs.}
\label{cone}
\end{figure}

Figure \ref{cone} displays a cone diagram for the A1489 survey projected along the right ascension. The characteristic large-scale structure is evident including thin filaments that delineate foreground and background voids. The foreground structure at $z \sim 0.176$ corresponds to the X-ray detected group 400dJ121259+272713 \citep{Burenin07}. 

The red points indicate members of A1489 (see Section \ref{sec:member}). The trumpet-shaped pattern associated with infall around the cluster is evident. The four colored stars mark the positions of the BCGs in redshift space. The galaxy denoted with a green star is closest to the dynamical center of the cluster. It is also the brightest object in the $r-$band (Figure \ref{cmd}).

%========================================
%========================================

\subsection{$\dn$ and Central Stellar Velocity Dispersion}\label{vdisp}

\begin{figure}[ht!]
\centering
\includegraphics[scale=0.38]{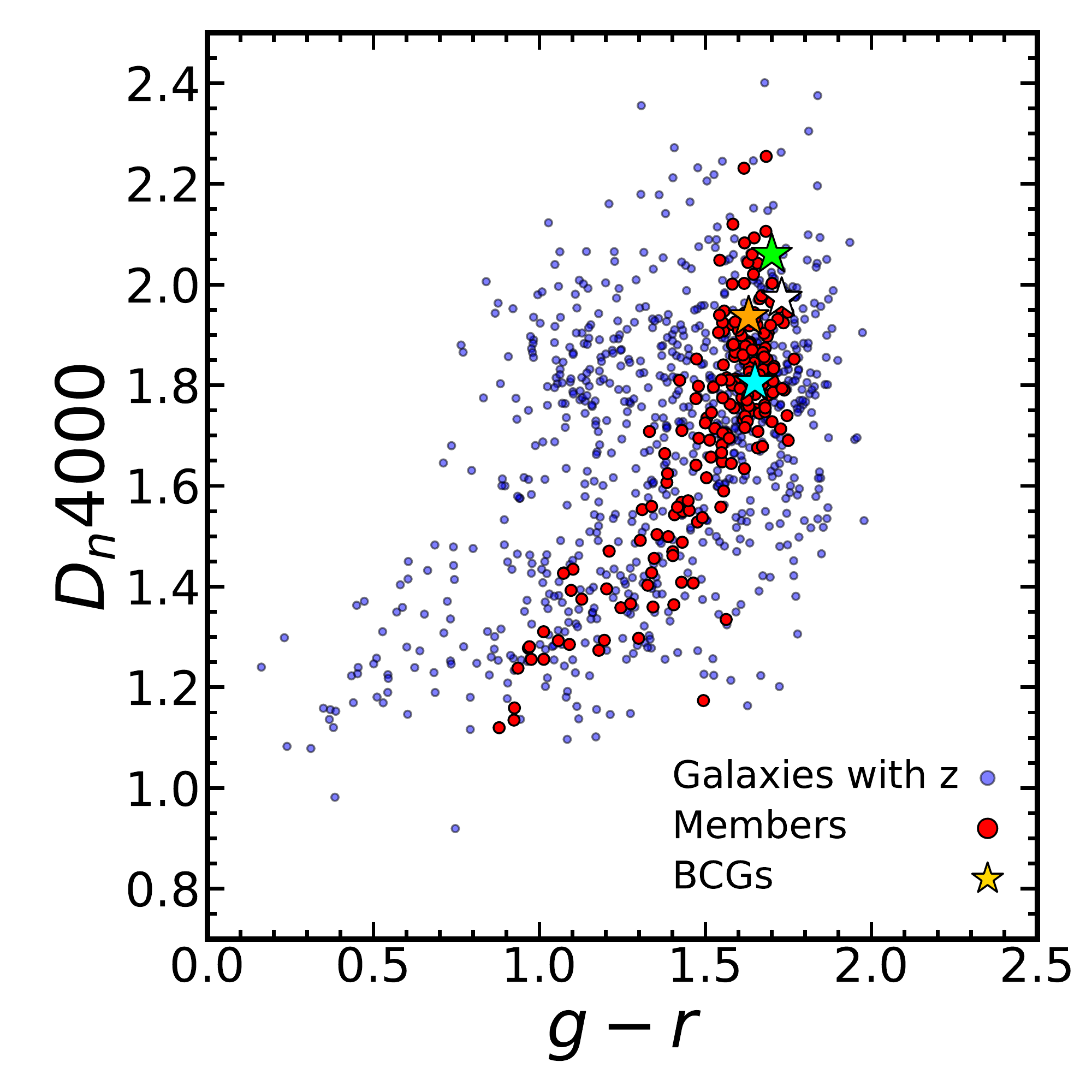}%{jsohn_figure/A1489_dn.pdf}
\caption{$\dn$ as a function of $g-r$ color. Blue circles show galaxies with a spectroscopic redshift and red circles display cluster members. The colored stars denote the four BCGs.}
\label{dn}
\end{figure}

The $\dn$ index is a stellar population age indicator (e.g., \citealp{Kauffmann03}). We follow the \citet{Balogh99} definition: $\dn$ is the flux ratio between $4000 - 4100~ {\AA}$ and $3850 - 3950~ {\AA}$: $\dn = F_{\lambda} (4000 - 4100) / F_{\lambda} (3850 - 3950)$. Figure \ref{dn} shows $\dn$ as a function of $(g-r)$ color. Galaxies with $(g-r) < 1$ have $\dn < 1.5$. As expected cluster members are predominantly red with $\dn \geq 1.5$. It is interesting to note that the 4 BCGs that delineate the cluster core have similar colors but span a significant range in $\dn$. The BCG coincident with the dynamical center (green star) has the largest $\dn$ suggesting that it contains the oldest central stellar population.

%========================================
%========================================

For galaxies with MMT/Hectospec spectra, we use the University of Lyon Spectroscopic analysis Software (ULySS, \citealp{Koleva09}) to compute the line-of-sight stellar velocity dispersion. ULySS compares the observed spectra with stellar population templates based on the PEGASE-HR code \citep{LeBorgne04} and the MILES stellar library \citep{SanchezBlazquez06}. To minimize the uncertainty in the velocity dispersion, we use the rest-frame spectral range $4100 - 5500$ {\rm \AA}~ for deriving the stellar velocity dispersion. A total of 70\% of the quiescent galaxies ($\dn > 1.5$) have a measured velocity dispersion.

For 203 galaxies with only an SDSS/BOSS spectrum, we collect the stellar velocity dispersion from the Portsmouth data reduction. This reduction \citep{Thomas13} measures the velocity dispersion by applying the Penalized Pixel-Fitting (pPXF) code \citep{Cappellari04}.

We apply an aperture correction to correct all of the velocity dispersion measurements to a fixed physical aperture of 3 kpc in the galaxy rest frame. The aperture correction is $\sigma_{A} / \sigma_{B} = (3~{\rm kpc} / R_{Hecto, SDSS})^{\beta}$, where $\sigma_{A}$ is the stellar velocity dispersion within a 3 kpc aperture and $\sigma_{B}$ is the measured dispersion within $R_{Hecto < SDSS}$, the appropriate fiber aperture in the galaxy rest frame for either Hectospec or SDSS \citep{Zahid17, Sohn17a}. \citet{Sohn17a} derive the coefficient $\beta = -0.054 \pm 0.005$ that we use here.

\subsection{Stellar Masses}\label{smass}

The strong lensing model of \citet{Zitrin20} for A1489 is based on a light-traces-mass approach. The model is based on the luminosities of probable cluster members  on  the photometric red sequence. Additional proxies for the masses of individual cluster members include the stellar mass and the central stellar velocity dispersion. Stellar masses have the obvious advantage that they correct for the population differences indicated by the range of $\dn$ values for cluster members (Figure \ref{dn}). 

For consistency with previous MMT redshift surveys (e.g., \citealp{Geller14, Zahid16}), we calculate stellar masses based on SDSS $ugriz$ model magnitudes corrected for foreground extinction. We fit the observed spectral energy distribution (SED) with the Le PHARE fitting code \citep{Arnouts99, Ilbert06}. We use the stellar population synthesis models of \citep{Bruzual03} and we assume a universal Chabrier initial mass function (IMF, \citealp{Chabrier03}). We consider a suite of models with two metallicities and with exponentially declining star formation rates. The e-folding time for star formation ranges from 0.1 to 30 Gyr. Model SEDs include various extinction levels and stellar population ages. We explore the internal extinction range E(B$-$V) = 0$-$0.6 based on the \citet{Calzetti00} extinction law. The population age range is 0.01 to 13 Gyr. We normalize each SED to solar luminosity. The ratio between the observed and synthetic SED is the stellar mass. We take the median of the distribution of best fit stellar masses as the estimate of the stellar mass for a particular object.

In Section \ref{sec:vdf} we compare the stellar masses and central velocity dispersions of cluster members. We also derive the velocity dispersion function and compare it with other strong lensing clusters.

\section{The Members of A1489} \label{sec:member}

We use the caustic technique \citep{Diaferio97, Diaferio99, Serra13} to identify the members of A1489. The caustic technique was originally designed as a method for computing the mass profile of a cluster based on a dense redshift survey. \citet{Serra13} demonstrate that the caustic technique is an efficient method to identify cluster members out to a radius of $3\rchar$. 

\begin{figure}[ht!]
\centering
\includegraphics[scale=0.38]{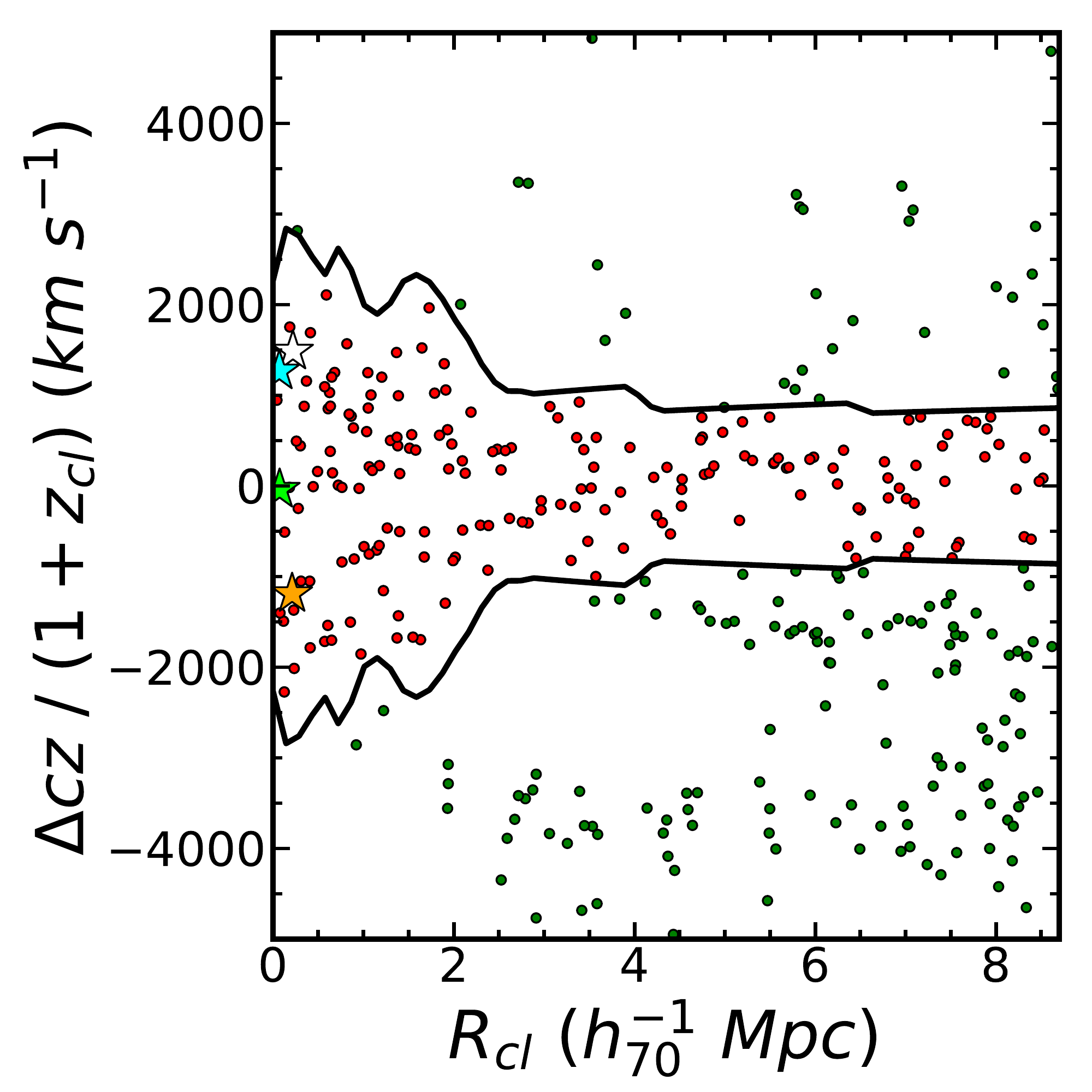}%{jsohn_figure/A1489_rv.pdf}
\caption{The R-v diagram for A1489. Red points indicate members within the caustic boundaries. Green points are non-members. Colored stars show the four BCGs associated with strong lensing arcs as in previous plots.}
\label{rv}
\end{figure}

To apply the caustic technique, we first plot the $R-v$ diagram for the cluster. This phase space diagram shows the rest-frame velocity of galaxies relative to the cluster center as a function of their clustercentric distance. Figure \ref{rv} shows the $R-v$ diagram for A1489. In agreement with \citet{Zitrin20} and with the X-ray observations, the brightest BCG (green star) is nearly coincident with the cluster mean redshift. The spectroscopy gives added weight to the identification of this object with the dynamical center of the cluster. 

%========================================
%========================================

In Figure \ref{rv}, the characteristic trumpet-shaped pattern defined by cluster members is evident. We identify the boundaries of the cluster (caustics) by computing the position where the phase space density changes steeply. \citet{Serra13} describe the detailed approach. We require that the caustic curves be symmetric relative to the cluster mean redshift. The solid curves in Figure \ref{rv} show the caustics. Red points in the Figure indicate the 195 cluster members; colored stars show the four BCGs. We discuss their remarkably different redshifts as a probable marker of the cluster history in Section \ref{sec:complex}.

\begin{figure} 
\centering
\includegraphics[scale=0.38]{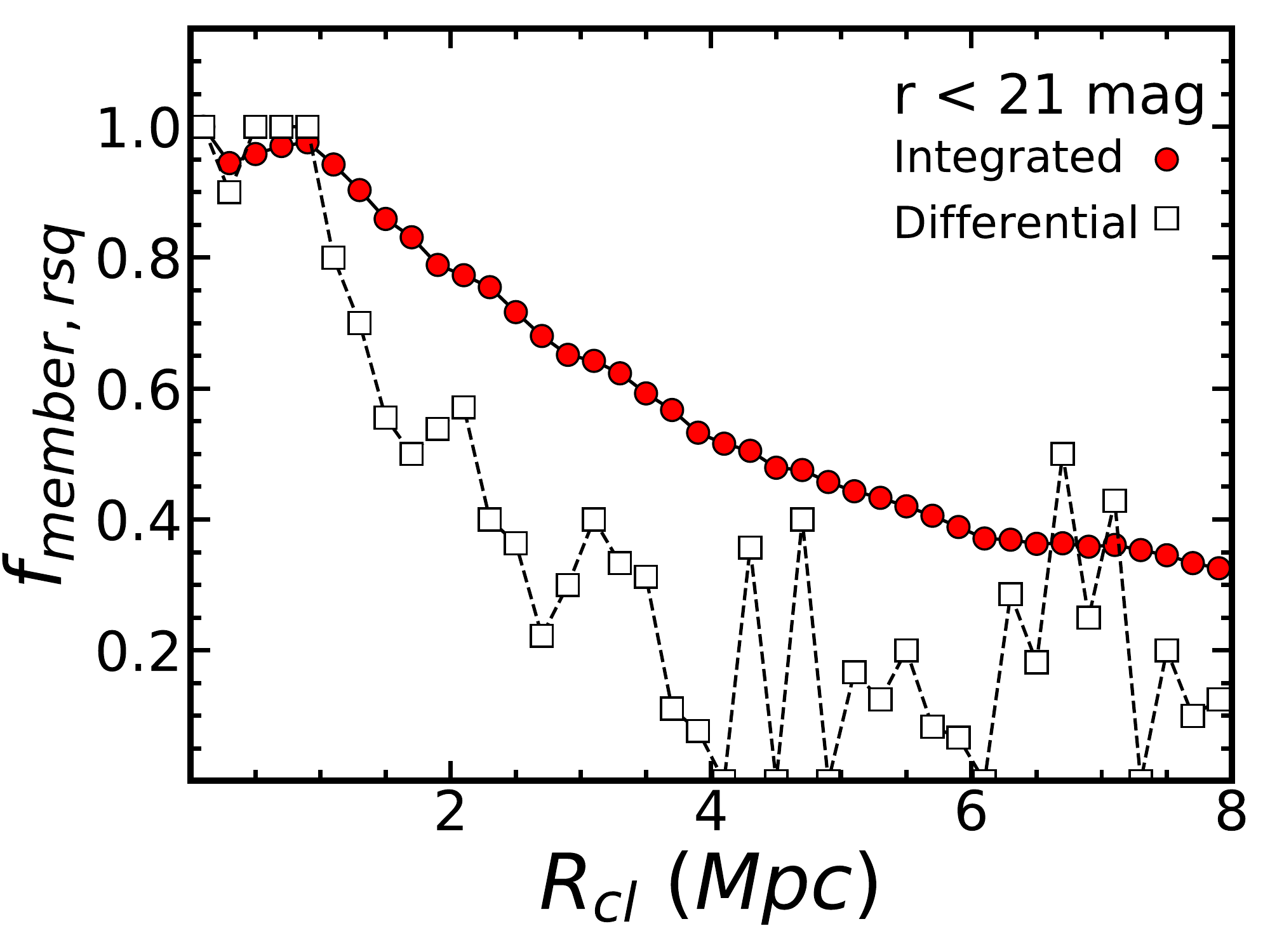}%{jsohn_figure/A1489_rsq_frac.pdf}
\caption{Spectroscopically identified member fraction among  red-sequence galaxies as a function of clustercentric distance. Red circles show the integrated fraction; white squares show the differential member fraction. }
\label{rsq}
\end{figure}

There are 195 cluster members within the caustic curves; 86 members lie within the characteristic radius $\rchar = 1.97$ Mpc. Figure \ref{rsq} shows the fraction of red sequence galaxies (Figure \ref{cmd}) that are cluster members as a function of clustercentric radius. Within $\sim 1$ Mpc (approximately 0.5$\rchar$), nearly all red sequence galaxies are cluster members. Within $\rchar$, more than 80\% of the members are on the red sequence, and the red fraction drops steeply at larger radius.

%========================================
%========================================

%========================================
\begin{figure} 
\centering
\includegraphics[scale=0.48]{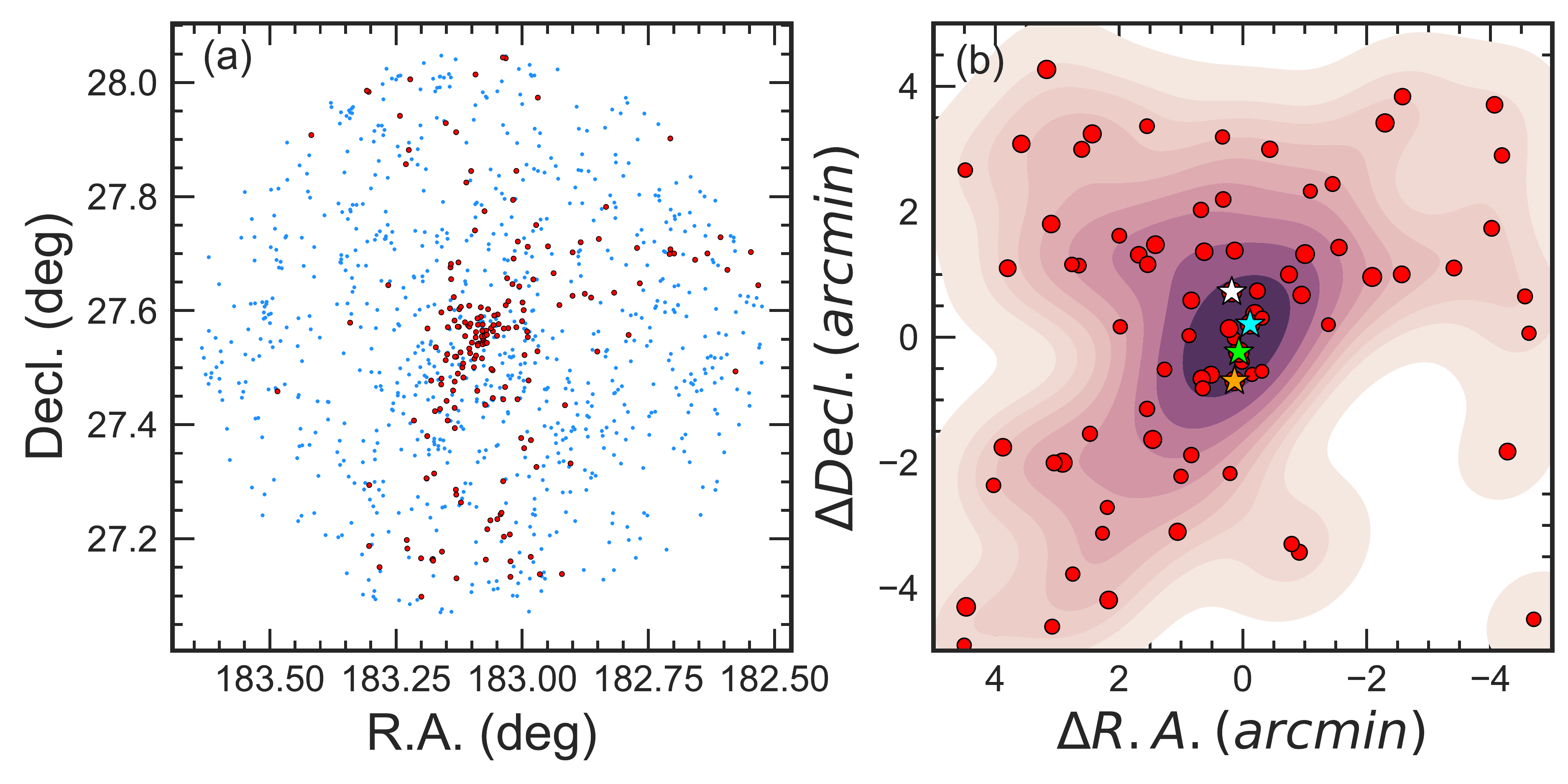}
%{jsohn_figure/A1489_spatial_combined.pdf}
\caption{(a) Spatial distribution of A1489 member galaxies (red circles). Blue circles show galaxies with spectroscopic redshifts. (b) Number density map of spectroscopic members of A1489. Red circles show individual member galaxies. Colored stars indicate the 4 BCGs associated with strong lensing arcs. } 
\label{spatial}
\end{figure}
%========================================

Figures \ref{spatial} explores the distribution of spectroscopically identified A1489 cluster members on the sky. Figure \ref{spatial} (a) shows the entire cluster redshift survey region. Red points show the cluster members identified with the caustic technique; blue points show non-members with spectroscopic redshifts. The elongated dense concentration of red points marks the core of the cluster imaged with HST (Figure \ref{hst}). At large radius the distribution of members is concentrated in a few regions. Given the broad distribution of non-members on the sky, the non-uniformity of the member distribution at large radius probably tracks the surrounding large-scale environment of the cluster. 

Figure \ref{spatial} (b) shows the central $10\arcmin \times 10\arcmin$ region of the cluster. This zoomed-in view highlights the positions of the four bright galaxies that dominate the core of the cluster.

\section{Velocity Dispersion Function of A1489}\label{sec:vdf}

The central stellar velocity dispersion of a quiescent galaxy is a good proxy for the dark matter subhalo dispersion \citep{Wake12, Bogdan15, Zahid16}. \cite{Zahid18} showed that the central stellar velocity dispersion is proportional to the dark matter velocity dispersion, which is proportional to the dark matter subhalo mass. Because the central stellar velocity dispersion is insensitive to the complex baryonic physics, measuring it is straightforward compared to other mass proxies including luminosity and stellar mass. These other mass proxies are also sensitive to systematic issues in the photometry for crowded fields.

\begin{figure}[ht!]
\centering
\includegraphics[scale=0.48]{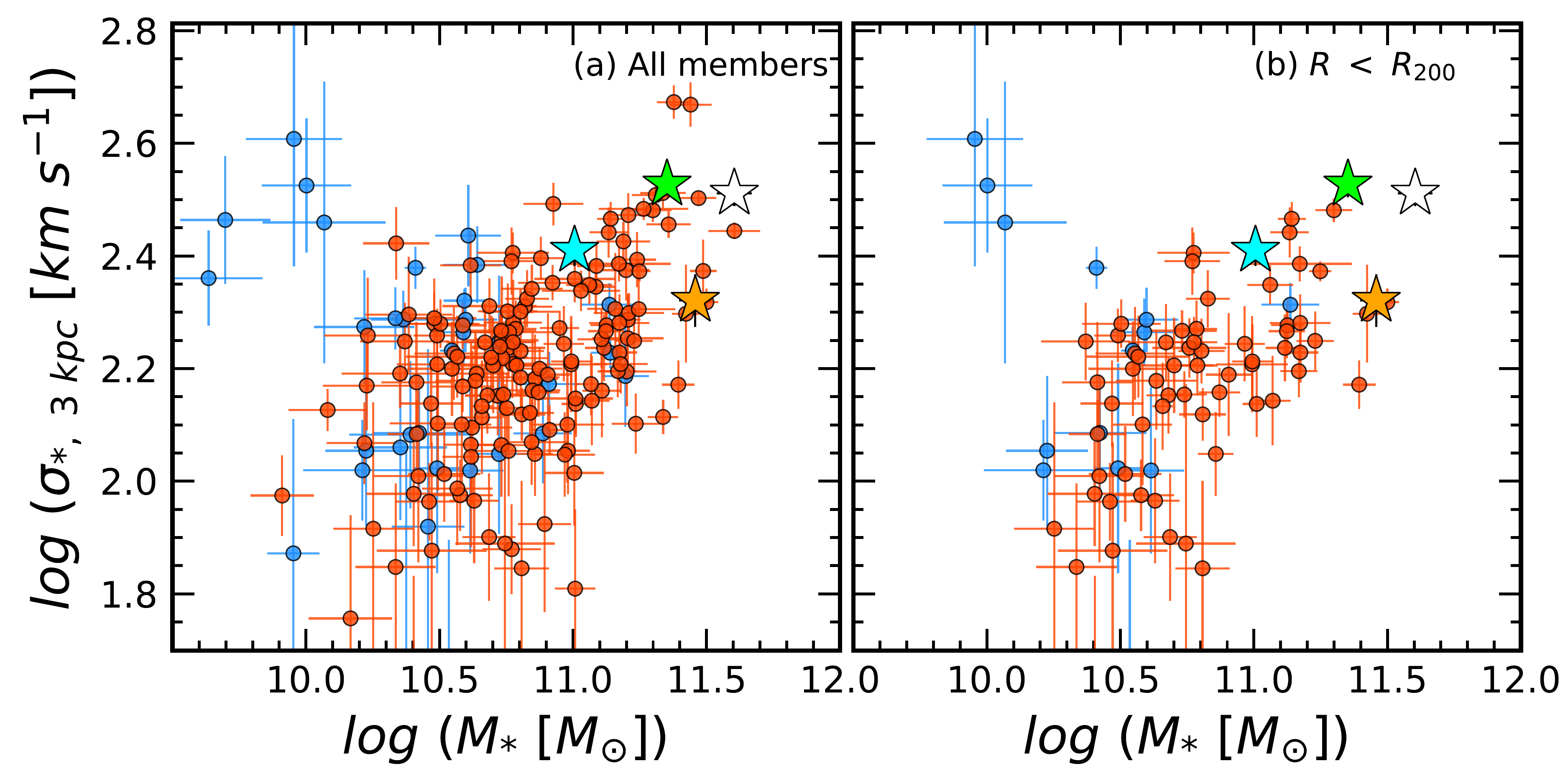}
%{jsohn_figure/A1489_sigma_mass_two.pdf}
\caption{Stellar velocity dispersion vs. stellar mass for (a) all cluster members and (b) members within $\rcl < \rchar$. Red and blue symbols indicate quiescent and star-forming objects. Stars mark the 4 BCGs with neighboring strong lensing arcs. }
\label{sigma_mass}
\end{figure}

In A1489, the observations include central stellar velocity dispersions for 188 spectroscopic members; 81 of them are within $\rchar$. Among these, 67 members are quiescent galaxies with $\dn > 1.5$. Figure \ref{sigma_mass} shows the stellar velocity dispersion as a function of stellar mass: (a) all cluster members, and (b) the members within $\rcl < \rchar$. Here, blue and red symbols indicate star-forming ($\dn \leq 1.5$) and quiescent ($\dn > 1.5$) cluster members. The two galaxy mass proxies are well correlated, but the scatter is large at every stellar mass. 

In Figure \ref{sigma_mass} (a), there are five galaxies with $\log \sigma_{*} > 2.3$ that lies above the general relation. These are star-forming objects with $\dn \leq 1.5$. There are also two quiescent objects with anomalously large $\sigma_{*}$. All of these objects including the two with large D$_n$4000 have emission lines in their spectra suggesting that a significant rotational component may contaminate the central velocity dispersion. 

Figure \ref{sigma_mass} (b) displays the tighter relation for members with $\rcl < \rchar$. Two of the BCGs including the BCG identified with the cluster center (green star) are among the objects with the largest central velocity dispersions.  

%========================================
%========================================

The velocity dispersion function (VDF) counts the number of cluster members as a function of stellar velocity dispersion. The VDF is a statistical tool that probes the subhalo mass distribution in a cluster in analogy with the luminosity or stellar mass functions. Cluster VDFs are measured for a small number of clusters: two local massive clusters (Coma and A2029, \citealp{Sohn17a}) and 9 strong lensing clusters at $0.18 < z < 0.29$ \citep{Sohn20b}. Here, we derive the A1489 VDF for the 67 quiescent member galaxies and compare it with other strong lensing systems that show an excess over the field at high central velocity dispersion.

The velocity dispersion function counts the number of  cluster member galaxies as a function of velocity dispersion. Although the idea is simple, measuring the VDF is not straightforward because of the inevitable incompleteness of the survey. For example, our survey for A1489 is only 50\% complete to $M_{r} \sim -21.5$, roughly corresponding to $\sigma = 150~\kms$. In order words, the raw number count at $\sigma = 150~\kms$ only represents 50\% of the member galaxies with similar velocity dispersion. Furthermore, there are some member galaxies without velocity dispersion measurements due to the low S/N of the spectra. 

We derive the A1489 VDF based on the method described in \citet{Sohn20b} (see their Section 3 for details). Here, we briefly review the process. We note that we derive the VDF for the members within $\rcl < \rchar$. We first count the number of photometric galaxies in $M_{r}$ and $\rcl$ bins. We compute the absolute magnitude of galaxies based on the cluster mean redshift. We then estimate the number of missing members ($N_{missing}$). We use the empirical member fraction as a function of $M_{r}$, $\dn$, and $\rcl$ derived from the HeCS-omnibus sample, a spectroscopic compilation of 227 clusters at $z < 0.3$ \citep{Sohn20a}. Based on this empirical member fraction, we estimate the number of missing spectroscopic members in each $M_{r}$ and $\dn$ bin. Here, we use $\dn$ bins because 1) the member fraction and 2) the velocity dispersion distribution depends on $\dn$ (see Figure 6 in \citealp{Sohn20b}).

We also count the number of spectroscopic members without velocity dispersion measurements ($N_{mem, no \sigma}$). The sum of $N_{missing}$ and $N_{mem, no \sigma}$ is the number of galaxies ($N_{cor}$) that need a statistical velocity dispersion estimate. For a statistical velocity dispersion estimate, we use the velocity dispersion distribution as a function of $M_{r}$ and $\dn$ derived from the HeCS-omnibus sample. Based on this distribution, we randomly draw $N_{cor}$ velocity dispersions. Finally, we construct the velocity dispersion function. We repeat these processes 1000 times and take the mean as the final VDF.  

%========================================
\begin{figure}[ht!]
\centering
\includegraphics[scale=0.48]{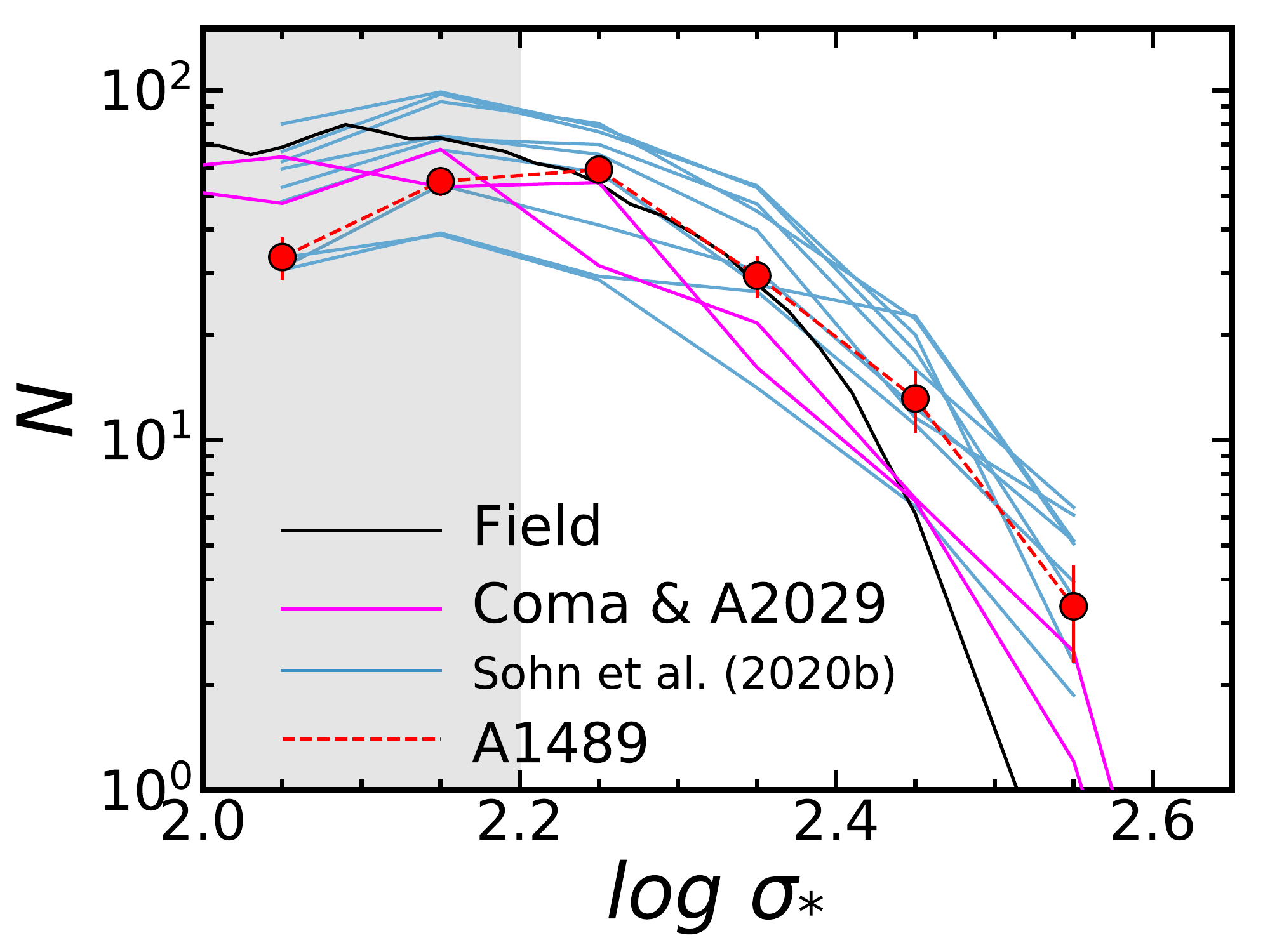}%{jsohn_figure/A1489_vdf.pdf}
\caption{Velocity dispersion function for A1489 quiescent members with $\rcl < \rchar$ (red circles and  dashed line). The blue lines show the velocity dispersion functions of 9 strong lensing clusters \citep{Sohn20b}. The magenta lines display the VDF of two local massive clusters: Coma and A2029 \citep{Sohn17a}. The black line indicates the field velocity dispersion function derived for SDSS quiescent galaxies \citep{Sohn17b}. In the shaded region, the velocity dispersion completeness of the cluster member survey is $< 50\%$.}
\label{vdf}
\end{figure}
%========================================

Figure \ref{vdf} shows the A1489 VDF (red symbols). For comparison, we plot the field VDF derived from the quiescent SDSS galaxies at $z < 0.1$ \citep{Sohn17b}. We normalize the field VDF at $\log \sigma = 2.2$. 

%We compare the VDF with those derived for two samples of clusters in a similar $\rchar$ and $\mchar$ range. 
The A1489 VDF shows a clear excess relative to the field at $\log \sigma > 2.4$. This excess is consistent with the excess in two local massive clusters, Coma and A2029 (magenta lines in Figure \ref{vdf}, \citealp{Sohn17a}). The blue lines show the VDFs measured from 9 strong lensing clusters at $0.18 < z < 0.29$ \citep{Sohn20b}. These cluster VDFs also show an excess at $\log \sigma > 2.4$. The normalization of the A1489 VDF is within the range spanned by these massive systems.

\section{The Dynamical Mass of A1489} \label{sec:dmass}

For a cluster that is sufficiently well-sampled in redshift space, the caustic technique provides a mass profile to large radius that is independent of any equilibrium assumption \citep{Diaferio97,Diaferio99,Serra11}. To obtain the mass profile, we identify the caustic curves (Figure \ref{rv}) with the escape velocity from the cluster. 

The vertical separation between the upper and lower caustics at radius $R$ is the caustic amplitude, $\mathcal{A}(R)$. The square of the caustic amplitude, $\mathcal{A}^2(R)$ estimates the square of the escape velocity, ${v_{\rm esc, los}^2}(R)$. With a filling factor $\mathcal{F}_\beta$ to account for cluster velocity anisotropy, the cluster mass profile is
\begin{equation}\label{eq:mass-prof}
G M(R) = \mathcal{F}_\beta \int_0^R \mathcal{A}^2(r)\,dr,
\end{equation}
where $G$ is the gravitational constant and $\mathcal{F}_\beta = 0.5$ is the filling factor (for further details see \citealp{Diaferio97,Diaferio99,Serra11}). 

%========================================
\begin{figure}[ht!]
\centering
\includegraphics[scale=0.48]{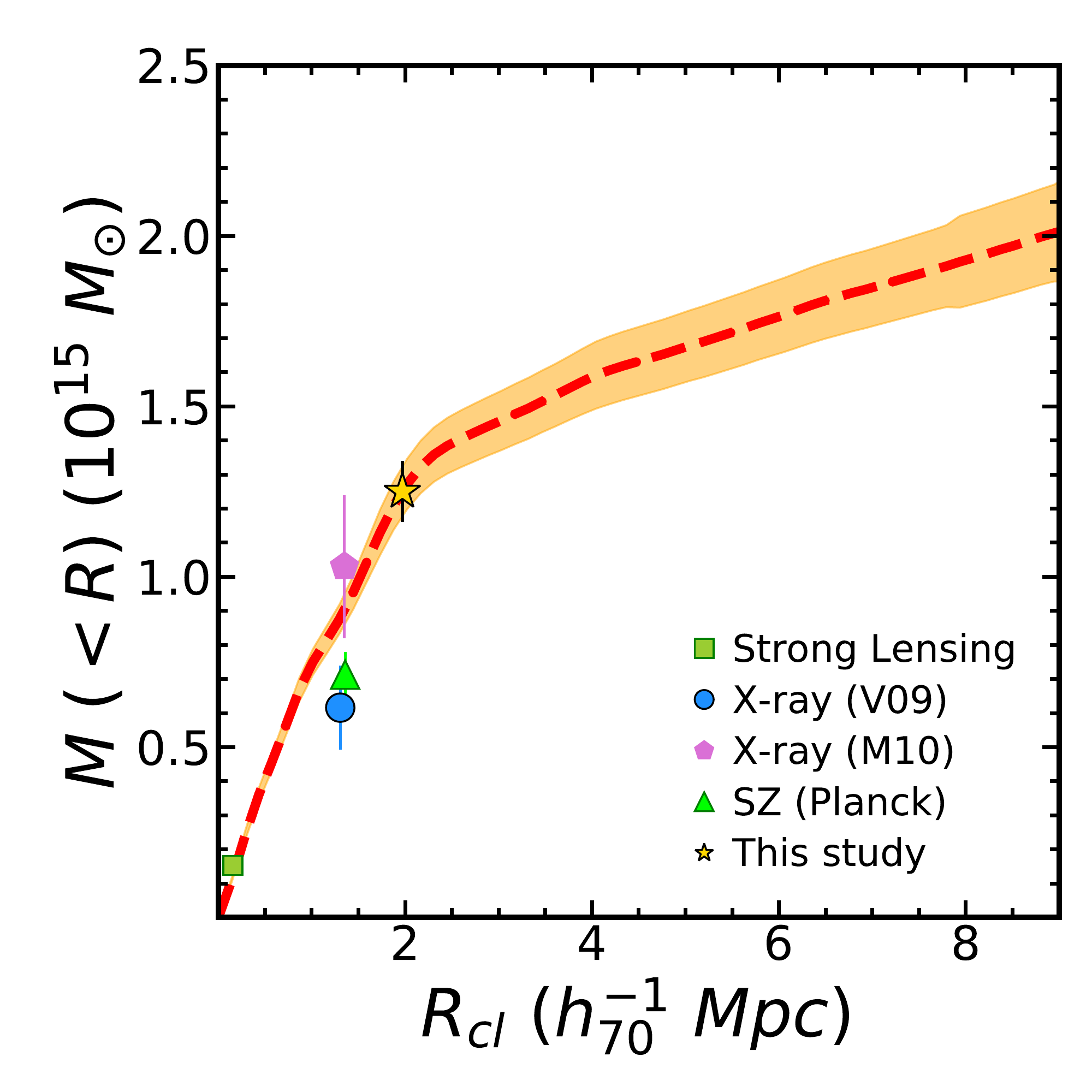}%{jsohn_figure/A1489_mprof.pdf}
\caption{Caustic mass profile of A1489 as a function of clustercentric distance (red dashed line and shaded region). The yellow star marks the $\rchar$ and $\mchar$ based on the caustic mass profile. The green square shows the mass estimate from strong lensing \citep{Zitrin20}. The blue circle and purple pentagon display the X-ray mass estimates from \citet{Vikhlinin09} and \citet{Mantz10}, respectively. The green triangle shows the SZ mass estimate \citep{Planck14}. These X-ray and SZ masses are measured within $R_{500}$. }
\label{mprof}
\end{figure}
%========================================

Figure \ref{mprof} shows the mass profile (dashed solid curve) based on the caustics for A1489. Table \ref{tab:a1489} lists $\rchar$, $\mchar$, and the line-of-sight velocity dispersion within $\rchar$. The yellow star in Figure \ref{mprof} indicates $\rchar$ and $\mchar$. 

The profile agrees with the strong lensing mass estimate \citep{Zitrin20} at small radius (green square in Figure \ref{mprof}). We note that like the caustic mass estimate, the strong lensing estimate is independent of assumptions about equilibrium.
 
\citet{Mantz10} use {\it Chandra} data to derive R$_{500} = 1.35 \pm 0.10$ Mpc in agreement with the $\rchar$ we derive from the caustics for an NFW mass profile  with concentration $c$=5 typical of massive clusters (for which $\rchar=1.5R_{500}$). Their mass estimate based on the gas mass, M$_{500} = (10.3 \pm 2.1) \times 10^{14}~\msol$ also agrees well with the caustic mass (purple pentagon, Figure \ref{mprof}). \citet{Vikhlinin09} derived a smaller mass (blue circle, Figure\ref{mprof}) from the same {\it Chandra} data. The reason for the differing results is unclear. 

The SZ signal of A1489 is clearly detected in {\em Planck} observations. Figure \ref{mprof} shows the mass derived from the SZ decrement, $Y_{SZ}$ \citep{Planck16} (green triangle). The mass calibration is based on {\em XMM-Newton} observations assuming hydrostatic equilibrium. The caustic mass profile lies above this mass estimate. 

\citet{Aguado21} suggest that SZ derived masses often underestimate the dynamical mass by $\sim 20\%$. \citet{Richard10} point out a possibly similar bias where the ratio of strong lensing to X-ray mass is 1.3 for unrelaxed systems. We discuss the unrelaxed nature of A1489 further in Section \ref{sec:complex}. These systematic offsets might account for the lower X-ray and SZ masses in Figure \ref{mprof}.

Figure \ref{mprof} highlights the importance of mass estimates that are independent of the cluster dynamical state. The cluster Cl0024+1654 supports this point. Cl0024+1654 is probably unrelaxed \citep{Czoske02}; the caustic and lensing mass estimates agree with each other but exceed the X-ray estimate \citep{Diaferio05}. A measured weak lensing profile extending to large radius would be the most important and informative test of the caustic mass profile for A1489. Extension of the caustic profile to even larger radius would also provide limits on the accretion rate and thus the future of A1489 \citep{Pizzardo21}.

The caustic estimates of $\rchar$ and $\mchar$ (Table \ref{tab:a1489}) for A1489 are in the range derived for other strong lensing systems with comparable or larger Einstein radii. \citet{Umetsu20} (Section 6.2) consider four superlensing clusters at a median redshift $z = 0.3$ and with Einstein radii $\Theta_E \geq 30\arcsec$. The median mass determined from lensing is  $\mchar = 13.4 \pm 0.9 \times 10^{14}~\msol$. \citet{Richard10} analyze 20 strong lensing clusters at a median $z = 0.2$ chosen on the basis of X-ray luminosity. The four clusters in their sample with large $\Theta_E$ include two of the systems in the study by \citet{Umetsu11} and \citet{ Umetsu20}. For the four large lenses (A1835, A1689, A1703, and A773) in the study by \citet{Richard10}, the $\rchar$ ranges from $1.99$ to $ 2.06$ Mpc and $\mchar$ ranges from $11.2$ to $ 13.2 \times 10^{14}~\msol$ \citep{Sohn20b}. In contrast with \citet{Umetsu20}, \citet{Sohn20b} use the caustic technique to determine $\rchar$ and $\mchar$. These four clusters are included in the comparison of cluster velocity dispersion functions (Figure \ref{vdf}).

%========================================
\section {Discussion}
%========================================

\subsection {A Complex Unrelaxed Cluster}\label{sec:complex}

\begin{figure}[ht!]
\centering
\includegraphics[scale=0.38]{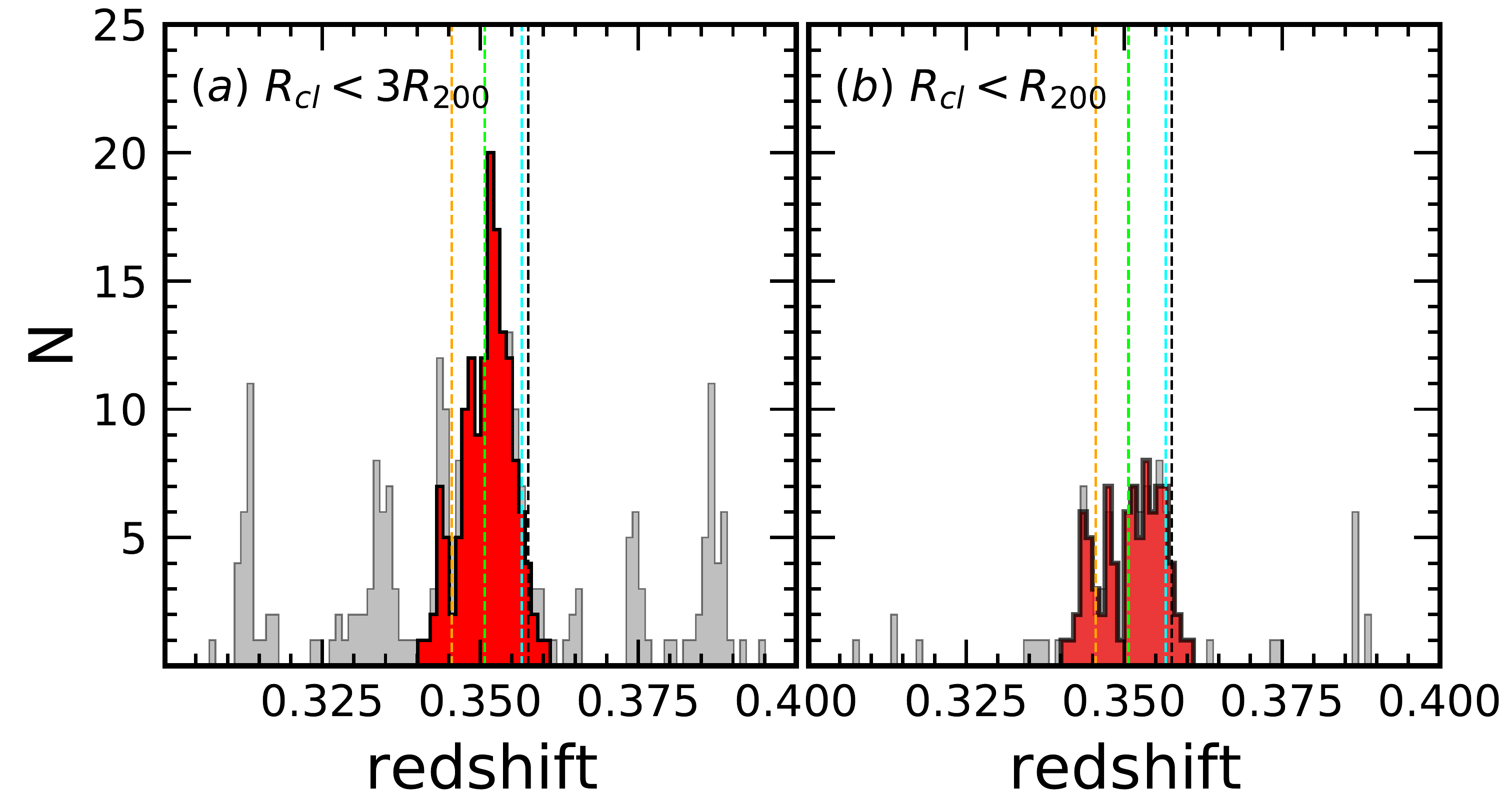}
%{jsohn_figure/A1489_dvhist_two.pdf}
\caption{The rest-frame line-of-sight velocity difference of galaxies with respect to the A1489 cluster center: (a) galaxies within $\rcl < 3\rchar$ and (b) within $\rcl < \rchar$. Gray histograms show all galaxies with spectroscopic redshifts, and the red histograms display cluster members. The vertical lines indicate the redshifts of four BCGs. }
\label{zhist}
\end{figure}

Figure \ref{zhist} shows rest-frame velocity histograms from members (red histograms) within $3\rchar$ (left panel) and within $\rchar$ (right). The colored arrows show the rest-frame velocities of the BCGs denoted with colored stars in previous figures. 

The BCG denoted with a green arrow is close to the peak of the histogram for the cluster within $3\rchar$ (left panel). In fact, its rest frame velocity is $-39~\kms$ relative to the overall cluster mean (Table \ref{tab:bcg}). This BCG is coincident with both the strong lensing cluster center and the X-ray center \citep{Zitrin20}. A Shapiro-Wilk test \citep{Shapiro65} of the Gaussianity of the full relative line-of-sight radial velocity distribution gives a $p = 0.078$ suggesting that the Gaussian distribution cannot be rejected.

In the zoomed in version of the histogram (right) showing the radial velocity distribution within a projected radius of $\rchar$, the central BCG coincides with a possible subsidiary peak. The redshifts of the other three BCGs have rest frame velocities $\gtrsim 1000~ \kms$ (Table \ref{tab:bcg}) relative to the cluster mean. These large velocity offsets are indicative of substructure in a complex unrelaxed cluster. These three BCGs appear to be associated with structure in the histogram. In this case, the Shapiro-Wilk test gives $p = 0.008$ rejecting the Gaussian null hypothesis and confirming the visual impression. Although there are 86 members in this histogram, the sampling is inadequate to decompose the probable velocity substructure within the system.

Other strong lensing systems with large Einstein radii, A1703, A1689, A1835, and A773, are included in Figure \ref{vdf}. Their median line-of-sight velocity dispersion within $\rchar$, $  1178 \pm 24~\kms$, \citep{Sohn20b} is essentially identical to the value for A1489 (Table \ref{tab:a1489}).

%========================================
%========================================

\begin{figure}[ht!]
\centering
\includegraphics[scale=0.48]{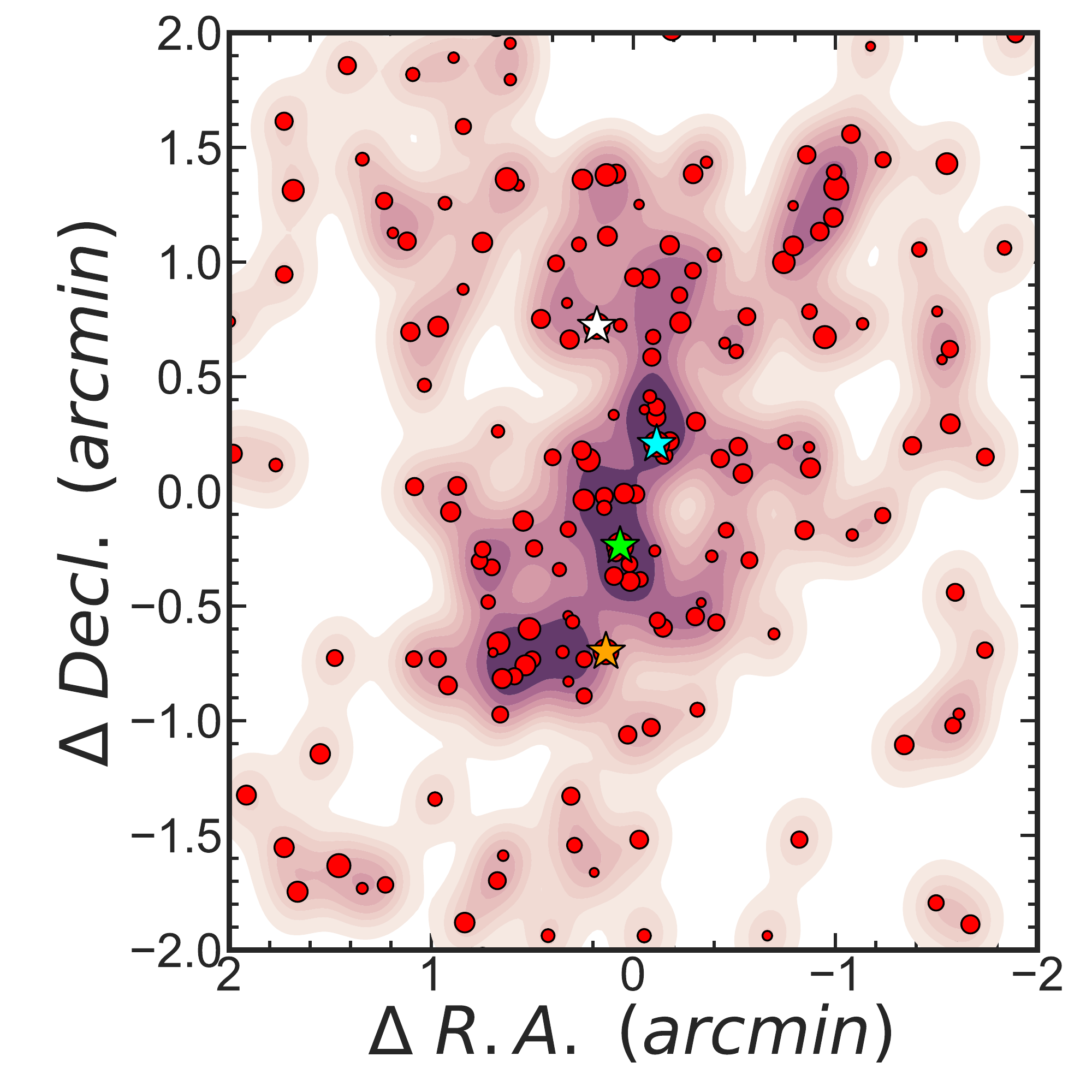}
%{jsohn_figure/A1489_zoom_spatial_contour_rsq.pdf}
\caption{Number density map of red-sequence galaxies (regardless of existing spectroscopic measurements) with $r < 23$ mag. The symbol sizes depend on the $r-$band magnitudes of the galaxies. Colored stars show the 4 BCGs. }
\label{zoom_rsq}
\end{figure}

Figure \ref{zoom_rsq} highlights possible substructure projected on the sky in the central region of the cluster. The contours show the surface number density of red sequence galaxies with $r-$band magnitudes $r < 23$. Most of these galaxies are likely cluster members. Three peaks in the surface number density appear to be associated with three of the BCGs. This apparent association requires confirmation with a deeper, more complete redshift survey that could resolve the structure cleanly in redshift space. Like the redshift histogram within $\rchar$, the multi-peaked surface number density suggests a complex system.

%========================================
%========================================

On the basis of N-body/hydrodynamical simulations, \citet{Meneghetti10} demonstrate that strong lensing clusters are dynamically more active than the typical system. They also note that the projected mass maps are elongated in the cluster core. They argue that this elongation reflects substructure. \citet{Richard10} find a correlation between substructure and the luminosity gap between the BCG and lower ranked cluster members. A1489 has an elongated core along with BCGs of comparable luminosity. These properties, together with the apparent complexity in the radial velocity histogram, suggest that A1489, like several other large lenses, is dynamically active.

\subsection{Central Stellar Velocity Dispersions and Strong Lensing Models} \label{sec:lensing}

A critical component of strong lens modeling is the separation of the mass associated with the halos of individual galaxies from the extended dark matter cluster halo. \citet{Monna15} introduced the use of stellar velocity dispersions of individual cluster members as a route toward more robust separations of the member galaxy and cluster halos. They demonstrate that use of the central stellar velocity dispersion breaks inherent degeneracies in strong lens modeling. For the cluster A383, use of velocity dispersions of member galaxies improves constraints on the truncation radii dark matter halos associated with cluster members by $\sim 50$\%, reducing the uncertainty to a few kpc out of a few tens of kpc. Concurrent constraints on the mass of the cluster halo improve by $\sim 10$\%. Application to other systems underscores the enhanced lens models enabled by the use of central velocity dispersions (e.g., \citealp{Monna17, Bergamini19, Granata21, Acebron21}).

In the crowded cluster environment, stellar velocity dispersions, in spite of the possible contamination by rotational motion, have an advantage over photometric measures because the effects of overlapping objects are minimized. Use of this set of dispersions will enhance constraints on the strong lensing models. The dataset for A1489 is larger than the one for A383 \citep{Monna15} but smaller than the recent sample used by \citet{Granata21} to analyze the cluster Abell S1063. A more complete sample of high dispersion objects and extension of the sample of dispersions to fainter cluster members would provide even stronger constraints. 

\section{Conclusion} \label{sec:conclusion}

A1489, a strong lensing cluster, is a target for upcoming JWST observations. Unlike several other clusters with large Einstein radii, A1489 was selected for HST observations on the basis of its unusual redMaPPer richness \citep{Rykoff14,Rykoff16} and predicted lensing strength. The Einstein radius is $\Theta_E = 32 \pm 3 \arcsec~$ for a source redshift $z_{S}$ = 2 \citep{Zitrin20}.

To complement the existing X-ray, SZ and HST observations, we provide a spectroscopic survey of A1489. The spectroscopy yields redshifts for 195 cluster members; 86 of these members are within the fiducial radius $\rchar$. The spectroscopy also provides central stellar velocity dispersions for 188 cluster members.

By applying the caustic technique \citep{Diaferio97,Diaferio99,Serra11} to the redshift survey we derive $\rchar$, $\mchar$, and the central line-of-sight velocity dispersion for A1489 (Table \ref{tab:a1489}). Agreement between the strong lensing and caustic masses emphasizes the importance of estimates that do not require the assumption of dynamical equilibrium. The parameters we derive are remarkably similar to the parameters that characterize other strong lensing clusters with similar Einstein radii.

We demonstrate that, like other relatively large $\Theta_{E}$ clusters, A1489 is probably unrelaxed. It contains four BCGs with remarkably different redshifts. One of the BCGs is coincident with the cluster dynamical center; the others have peculiar velocities of $\gtrsim 1000~\kms$. A denser redshift survey would probably resolve the complex structure indicated by the rest-frame velocity histogram and by the distribution of faint red sequence galaxies in the cluster core.

We also emphasize the power of the central stellar velocity dispersions of cluster members for refining the cluster properties. The velocity dispersion function for the members of A1489 has an excess at dispersions $\gtrsim 250~\kms$ similar to other massive clusters, including several that are strong lenses with large Einstein radii. Furthermore, the velocity dispersions along with larger future samples should tighten constraints on strong lensing models of the cluster. 

\acknowledgements
J.S. is supported by a CfA Fellowship. M.J.G. acknowledges the Smithsonian Institution for support. A.D. acknowledges partial support from  the INFN grant InDark and the Departments of Excellence grant L.232/2016 of the Italian Ministry of Education, University and Research (MIUR). 
This research has made use of NASA’s Astrophysics Data System Bibliographic Services.

Funding for the SDSS-IV has been provided by the Alfred P. Sloan Foundation, the U.S. Department of Energy Office of Science, and the Participating Institutions. SDSS-IV acknowledges support and resources from the Center for High Performance Computing at the University of Utah. The SDSS website is www.sdss.org. SDSS-IV is managed by the Astrophysical Research Consortium for the Participating Institutions of the SDSS Collaboration including the Brazilian Participation Group, the Carnegie Institution for Science, Carnegie Mellon University, Center for Astrophysics | Harvard and Smithsonian, the Chilean Participation Group, the French Participation Group, Instituto de Astrofísica de Canarias, Johns Hopkins University, Kavli Institute for the Physics and Mathematics of the Universe (IPMU)/University of Tokyo, the Korean Participation Group, Lawrence Berkeley National Laboratory, Leibniz Institut für Astrophysik Potsdam (AIP), Max-Planck-Institut für Astronomie (MPIA Heidelberg), Max- Planck-Institut für Astrophysik (MPA Garching), Max-Planck- Institut für Extraterrestrische Physik (MPE), National Astro- nomical Observatories of China, New Mexico State University, New York University, University of Notre Dame, Observatário Nacional/MCTI, Ohio State University, Pennsylvania State University, Shanghai Astronomical Observatory, United King- dom Participation Group, Universidad Nacional Autónoma de México, University of Arizona, University of Colorado Boulder, University of Oxford, University of Portsmouth, University of Utah, University of Virginia, University of Washington, University of Wisconsin, Vanderbilt University, and Yale University.

\bibliographystyle{aasjournal}
\bibliography{ms}

\end{document}